\documentclass{nature}

\usepackage{graphicx,amsmath,amssymb,geometry,pdflscape}

\makeatletter
\let\saved@includegraphics\includegraphics
\AtBeginDocument{\let\includegraphics\saved@includegraphics}
\renewenvironment*{figure}{\@float{figure}}{\end@float}
\makeatother

\usepackage{astjnlabbrev-nature} 

\title{A hidden population of high-redshift double quasars unveiled by astrometry}

\author{Yue Shen$^{1,2}$, Yu-Ching Chen$^{1}$, Hsiang-Chih Hwang$^3$, Xin Liu$^{1,2}$, Nadia Zakamska$^3$, Masamune Oguri$^{4,5}$, Jennifer I-Hsiu Li$^{1}$, Joseph Lazio$^6$, Peter Breiding$^{7}$}

\newcommand{\arcsec}{\mbox{\ensuremath{^{\prime\prime}}}}
\newcommand{\farcs}{\mbox{\ensuremath{.\!\!^{\prime\prime}}}}

\newcommand{\CIV}{C{\sevenrm IV}}

   \font\sevenrm=cmr7 scaled 1000

\usepackage{color,subfigure}

\begin{document}

\maketitle

\begin{affiliations}
 \item Department of Astronomy, University of Illinois at Urbana-Champaign, Urbana, IL 61801, USA
 \item National Center for Supercomputing Applications, University of Illinois at Urbana-Champaign, Urbana, IL 61801, USA
 \item{Department of Physics and Astronomy, Johns Hopkins University, Baltimore, MD 21218, USA}
 \item{Research Center for the Early Universe and Department of Physics, University of Tokyo, Japan}
 \item{Kavli Institute for the Physics and Mathematics of the Universe (Kavli IPMU), University of Tokyo, Japan}
 \item{Jet Propulsion Laboratory, California Institute of Technology, M/S 67-201, Pasadena, CA  91109, USA}
 \item{Department of Physics and Astronomy, West Virginia University, P.O. Box 6315, Morgantown, WV 26506, USA}
\end{affiliations}

\begin{abstract}

Galaxy mergers occur frequently in the early universe\cite{Duncan} and bring multiple supermassive black holes (SMBHs) into the nucleus, where they may eventually coalesce. Identifying post-merger-scale (i.e., $\lesssim$\,a few kpc) dual SMBHs is a critical pathway to understanding their dynamical evolution and successive mergers\cite{Begelman_1980}. While serendipitously discovering $\sim$\,kpc-scale dual SMBHs at $z<1$ is possible\cite{komossa03}, such systems are elusive at $z>2$, but critical to constraining the progenitors of SMBH mergers. The redshift $z\approx 2$ also marks the epoch of peak activity of luminous quasars\cite{Richards06}, hence probing this spatial regime at high redshift is of particular significance in understanding the evolution of quasars. However, given stringent resolution requirements, there is currently no confirmed $<10\,{\rm kpc}$ physical SMBH pair at $z>2$ [ref.~5-8]. Here we report two sub-arcsec double quasars at $z>2$ discovered from a targeted search with a novel astrometric technique, demonstrating a high success rate ($\gtrsim 50\%$) in this systematic approach. These high-redshift double quasars could be the long-sought kpc-scale dual SMBHs, or sub-arcsec gravitationally-lensed quasar images. One of these double quasars (at $z=2.95$) was spatially resolved with optical spectroscopy, and slightly favors the scenario of a physical quasar pair with a projected separation of 3.5\,kpc (0\farcs46). Follow-up observations of double quasars discovered by this targeted approach will be able to provide the first observational constraints on kpc-scale dual SMBHs at $z>2$. 

\end{abstract}

The principles of the astrometry-based technique are\cite{Liuy,vodka1}: (1) the unresolved source contains multiple components; (2) these components have unsynchronized flux variations that cause astrometric jitter in the photocenter of the system; (3) this photocenter jitter is detectable with astrometric measurements. Extragalactic targets that fulfill these requirements include unresolved physical pairs (or multiples) of quasars and small-scale gravitationally lensed quasars. In the case of physical pairs, the variability of each quasar is stochastic and uncorrelated\cite{MacLeod10}; in the case of gravitationally lensed quasars, the variability from each image is also unsynchronized due to lensing time delays\cite{Refsdal}. Assuming the pair of quasars have comparable mean fluxes and rms fluctuations, the relation between the expected astrometric jitter $\sigma_{\rm astro}$, the separation of the pair $D$, and the variability amplitude of the system is\cite{vodka1}: 
\begin{equation}\label{eqn:jitter}
\sigma_{\rm astro}\approx \frac{D}{2}\frac{\sqrt{\left<\Delta f^2\right>}}{\bar{f}}\ ,
\end{equation}
where $\bar{f}$ and $\sqrt{\left<\Delta f^2\right>}$ are the total mean and rms fluxes of the unresolved system. This relation can be generalized to non-equal-flux pairs, for which the expected astrometric jitter is reduced\cite{vodka1}, rendering this technique less efficient. Considering typical fractional rms fluctuations of quasars of $\sim 10\%$ [ref.~11], the expected astrometric jitter is $\sim 10$\,milli-arcsec for 0\farcs2 separations, corresponding to sub-kpc to kiloparsec projected separations at cosmological distances. 

Ground-based and seeing-limited imaging cannot easily resolve quasar pairs or lensed quasars with sub-arcsec separations, while diffraction-limited imaging with HST or ground-based adaptive optics cannot probe large volumes. However, the anticipated astrometric jitter from such systems is well within the reach of the Gaia astrometry satellite\cite{Gaia,Gaiadr2} covering the entire optical sky with roughly uniform photometric and astrometric sensitivities, enabling a novel targeted approach to systematically discover such systems with high efficiency. 

We design a targeted search for sub-arcsec pairs among luminous quasars that appear to be singles in seeing-limited surveys, focusing on the $z>2$ regime where mergers are expected to be more common. First, we select bright, spectroscopically confirmed quasars from the SDSS\cite{Shen_etal_2011,dr14q} at $z>2$. We require that Gaia only detects a single source within a $3$\arcsec\ radius of the SDSS position, hence the quasar is unresolved at the Gaia resolution of $\sim 0\farcs2-0\farcs8$ depending on the scanning scheme of Gaia. The redshift cut ensures that host galaxy emission is negligible and the optical light is dominated by the quasar light. At lower redshifts where the more extended host galaxy is detectable with Gaia, the astrometric solution may be compromised given the scanning directions of Gaia and the astrometric modeling\cite{Lindegren18,vodka1}. Then, we restrict to objects where Gaia detects significant astrometric excess noise\cite{Lindegren12} -- a proxy for the astrometric jitter signal expected from an unresolved quasar double. The full details of target selection are provided in Methods. Our selection of small-scale quasar pairs using Gaia is thus markedly different from using resolved multiple Gaia sources\cite{lemon17}, and can probe much smaller pair separations. 

This astrometric selection with Gaia resulted in 15 candidates. Four of them were randomly observed by a HST Snapshot program (see Methods and Extended Data Figure~1). We show the 2-band composite images of these four candidates in Figure~1 and Extended Data Figure~2. Two of them (J0749 and J0841) reveal two point-like cores separated by $\sim 0\farcs5$ with comparable fluxes and similar colors, with no obvious detection of extended features. In addition, the optical spectra from SDSS (Figure~2 and Extended Data Figure~3) enclosing all light within a $2-3$\arcsec\ diameter fiber rule out an obvious quasar-star superposition or (unrelated) projected quasar pairs. Therefore these two objects are strongly favored for physical quasar pairs or lensed quasars. Additional observations confirming their double-quasar origin are provided in Methods and Extended Data Figure~4. The third object, J0905, has two resolved point-like components with different colors and very small flux ratio in the blue band. Thus J0905 is most likely a quasar-star superposition that cannot be easily ruled out from the SDSS spectrum, as further reasoned in Methods. The remaining object, J1326, is a single point source under HST resolution of $\sim 0\farcs1$, indicating that Gaia astrometric excess noise is caused by pairs within 0\farcs1, or is due to unknown Gaia systematics. Extended Data Figure~5 shows the imaging decomposition results of the HST data.

We estimate the frequency of $z>2$ SDSS quasars that are apparent singles in seeing-limited surveys but in fact are doubles (or multiples) on sub-arcsec scales using the statistics from our targeted search with astrometry. Assuming the frequency is not a function of limiting magnitude, we had 15 targets selected with Gaia astrometric excess noise for HST follow-up, out of $\sim 11,000$ parent quasars, or $0.14\%$. This is a hard lower limit, since most of these quasars could be doubles at smaller separations and/or their variability amplitude is low, thus the astrometric jitter would be well below the Gaia astrometric excess noise cut that we imposed in target selection. Among the four randomly observed targets with HST, two are double quasars on sub-arcsec scales, or a success rate of $50\%$. This is again a lower limit, since the separation can be below HST resolution of $\sim 0.1$\arcsec\ in the fourth object. The combined lower limit of the frequency is thus $\sim 0.1\%$. Incidentally, this rate is consistent with the estimated rate of $<0.4\%$ for lensed quasars at $>1$\arcsec\ image separations\cite{Pindor}. We can also compare to the kpc-scale physical pair fraction extrapolated from small-scale quasar clustering measurements at $z<2$ [ref.~21], which suggested a pair fraction of $\sim 0.05\%$ (Methods). The anticipated frequency of kpc-scale physical quasar pairs may be substantially higher at $z>2$ (see Methods). 

To further illuminate the nature of the double quasars discovered by Gaia astrometry, we acquired optical slit spectroscopy for one of the double-cored targets (J084129.77+482548.5, hereafter J0841) with Gemini under excellent seeing conditions. The system is spatially resolved in the slit spectroscopy (see Methods and Extended Data Figure~6) at the locations of the two HST cores. The two spatially-resolved spectra are presented in Figure~2, which are rather similar quasar spectra. The flux ratio of the two cores averaged over the optical spectral range is $\sim 1.5$, consistent with that measured from HST imaging. 

While there are spectral differences in the two cores of J0841 (see Methods and Extended Data Figure~7), notably in the strengths of the broad emission lines, the possibility of lensing cannot be conclusively ruled out. We thus model the system as a gravitationally lensed quasar, where the lens galaxy is undetected in HST imaging (see Methods). There are only three other gravitationally lensed double quasars with $<0\farcs8$ separations known at $z>2.5$ [ref.~22-24], all of which were serendipitously discovered, unlike J0841 which is from a targeted search. The black line in the top panel of Figure~3 is the required lens velocity dispersion at a given lens redshift. We further use the HST non-detection to place $3\sigma$ upper-limits of the lens galaxy, indicated by the colored lines, which lie below the lens model constraint at $z<1.5$. This suggests that a lens galaxy at $z>1.5$ is not ruled out by HST non-detections. We caution that our detection limit estimates for extended sources may be overly optimistic, and more conservative non-detection limits will lead to less stringent constraints on the lensing hypothesis (see Methods for details).
 
In addition, we can estimate the lensing probability using the abundance of lens galaxies and impact parameters of the lensing system. The red line in the bottom panel of Figure~3 is the cumulative number of lens that could have produced a lensed quasar as in J0841. Taking into account the parent quasar sample, lensing magnification bias, and the probability of observing the lensed quasar in our follow-up, we estimate a total probability of $\sim 5\%$ that J0841 is a doubly lensed quasar. Based on the same statistical arguments, we can estimate the total number of sub-arcsec lensed quasars expected from our parent sample to be $\sim 2-3$, less than the number of double quasars anticipated from our systematic search (see Methods). However, given the uncertainties in the assumptions in our statistical arguments, the existing data are insufficient to rule out the possibility that the sub-arcsec double quasars discovered by astrometry are dominated by the lensed population. 

On the other hand, many observed high-redshift double quasars at $>1\arcsec$ are physically associated rather than lensed images\cite{Djorgovski,Kochanek,Hennawi10,Eftekharzadeh17}, and our initial results with this systematic approach can place some interesting constraints on the $<10$\,kpc binary supermassive black hole (SMBH) population at $z>2$ for the first time. Our targeted search with Gaia astrometry preferentially selects pairs with projected $\sim\,{\rm kpc}$ physical separations at $z>2$ and with comparable fluxes, i.e., typically from major mergers. If we assume all of the sub-arcsec pairs discovered with the astrometric technique are kpc-scale physical pairs on their way to coalesce, their frequency is $\sim 0.1\%$ among all optically-selected $z>2$ quasars. Since quasars light up stochastically, not all kpc-scale dual SMBHs are both quasars at the same time, and we will only observe a fraction of them as quasar doubles. The duty cycle of quasars is roughly 1 percent at $z>2$, based on recent SDSS quasar clustering measurements\cite{Eftekharzadeh15}, and we assume being in a kpc-scale binary does not alter the duty cycle. Thus the frequency of $\sim$\,kpc dual SMBHs among all optical quasars at $z>2$ would be $\sim 10\%$. The dynamical evolution of the binary SMBH from tens of kpc to $\sim 10$\,parsec is determined by dynamical friction\cite{Begelman_1980,Yu_2002}, although gas drag may not be negligible in high-redshift gas-rich mergers. With dynamical friction alone, in general the SMBH pair spends about a factor of ten less time\cite{Yu_2002} on hundreds of parsec than on kpc-scales probed by our observations. Thus by extrapolating our results on kpc-scales to scales a factor of ten smaller, we expect $\sim 1\%$ of $z>2$ optical quasars are sub-kpc dual SMBHs, and $\sim 0.01\%$ of them will appear as sub-kpc double quasars. 

An alternative scenario for J0841 is that it is a physical quasar pair, but not from the merger of two galaxies. High-redshift galaxies in formation are often clumpy\cite{Elmegreen}, with massive clumps separated by up to 10\,kpc. It is possible that these clumps, which constitute the bulk stellar mass of the galaxy, each contain a SMBH at the center. The formation of these galactic clumps through disk instabilities may have also facilitated gas inflow within each clump to feed onto seed SMBHs. In this scenario, the growth of the two SMBHs may be in-sync, resulting in rather similar spectral appearances and physical properties such as BH mass when witnessed as a close pair of quasars. However, deep infrared imaging and kinematics studies are required to test this scenario of multiple SMBH formation in clumpy high-redshift galaxies.

The novel technique based on the precision Gaia astrometry and physical causes of astrometric jitter in unresolved distant sources opens a new window to systematically discover and characterize sub-arcsec physical quasar pairs at high redshift, after the contamination from lensed quasars is properly quantified with detailed follow-up observations. This population of quasar pairs is expected to exist, but so far has been observationally elusive. The efficiency of this approach is demonstrated to be high based on follow-up observations with HST imaging and slit spectroscopy. The technique is most effective for pairs with comparable fluxes which produce the most significant astrometric jitter in the systemic photocenter\cite{vodka1}. Since spatially integrated spectroscopy can be used to rule out obvious quasar-star pairs, while galaxies are generally too faint to affect the photocenter at $z>2$, spatially resolved optical imaging provides strong evidence for the double quasar nature (either a physical pair or lensed images) to be confirmed by additional observations, as demonstrated here.

Further confirmation of quasar doubles selected with this technique but with separations below HST resolution can be achieved with JWST (with a factor of few better resolution in the optical than HST) and/or radio interferometry, as well as AO-assisted near-IR slit spectroscopy. In particular, near-IR spectroscopy will cover narrow emission lines of these high-redshift quasars to identify potential velocity offsets in the narrow-line emission, which can also signal physical quasar pairs without the need to spatially resolving the pair\cite{Liu10}. This targeted approach provides candidates with high purity, and enables efficient follow-up observations to confirm the nature of physical pairs or lensed images. In turn, it will lead to much improved statistics of high-redshift small-scale dual SMBHs, and a much better understanding of the kpc to sub-kpc environment of high-redshift quasars as well as the sub-arcsec lensed quasar population.


\begin{addendum}
 \item[Correspondence] Correspondence and requests for materials
should be addressed to Y.S. \\
(email: shenyue@illinois.edu).

 \item Support for the work of Y.S. was provided by an Alfred P. Sloan Research Fellowship and NSF grant AST-2009947. The work of M.O. was supported by World Premier International Research Center Initiative (WPI Initiative), MEXT, Japan, and JSPS KAKENHI Grant Number JP18K03693, JP20H00181, and JP20H05856. The NANOGrav project receives support from National Science Foundation (NSF) Physics Frontiers Center award number 1430284.  Part of this research was carried out at the Jet Propulsion Laboratory, California Institute of Technology, under a contract with the National Aeronautics and Space Administration. We thank John Blakeslee for granting us Gemini DDT time. Based on observations obtained at the international Gemini Observatory, a program of NSF’s NOIRLab, which is managed by the Association of Universities for Research in Astronomy (AURA) under a cooperative agreement with the National Science Foundation on behalf of the Gemini Observatory partnership: the National Science Foundation (United States), National Research Council (Canada), Agencia Nacional de Investigaci\'{o}n y Desarrollo (Chile), Ministerio de Ciencia, Tecnolog\'{i}a e Innovaci\'{o}n (Argentina), Minist\'{e}rio da Ci\^{e}ncia, Tecnologia, Inova\c{c}\~{o}es e Comunica\c{c}\~{o}es (Brazil), and Korea Astronomy and Space Science Institute (Republic of Korea). Based on observations made with the NASA/ESA Hubble Space Telescope, obtained at the Space Telescope Science Institute, which is operated by the Association of Universities for Research in Astronomy, Inc., under NASA contract NAS 5-26555. These observations are associated with program number GO-15900. The National Radio Astronomy Observatory is a facility of the National Science Foundation operated under cooperative agreement by Associated Universities, Inc. 
 
 \item[Author Contributions] Y.S. designed the project and led the analysis and manuscript writing; X.L. led the Gemini program and Y.C.C. reduced the Gemini spectra; H.C.H and N.Z. led the HST program; M.O. led the lensing analysis; J.I.L. performed the HST imaging analysis; X.L., Y.C.C., J.L. and P.B. contributed to the VLBA data; all authors contributed to the science interpretation. 
 
 \item[Competing Interests] The authors declare that they have no
competing financial interests.

\end{addendum}

\begin{figure}
\centering
\includegraphics[width=0.8\textwidth]{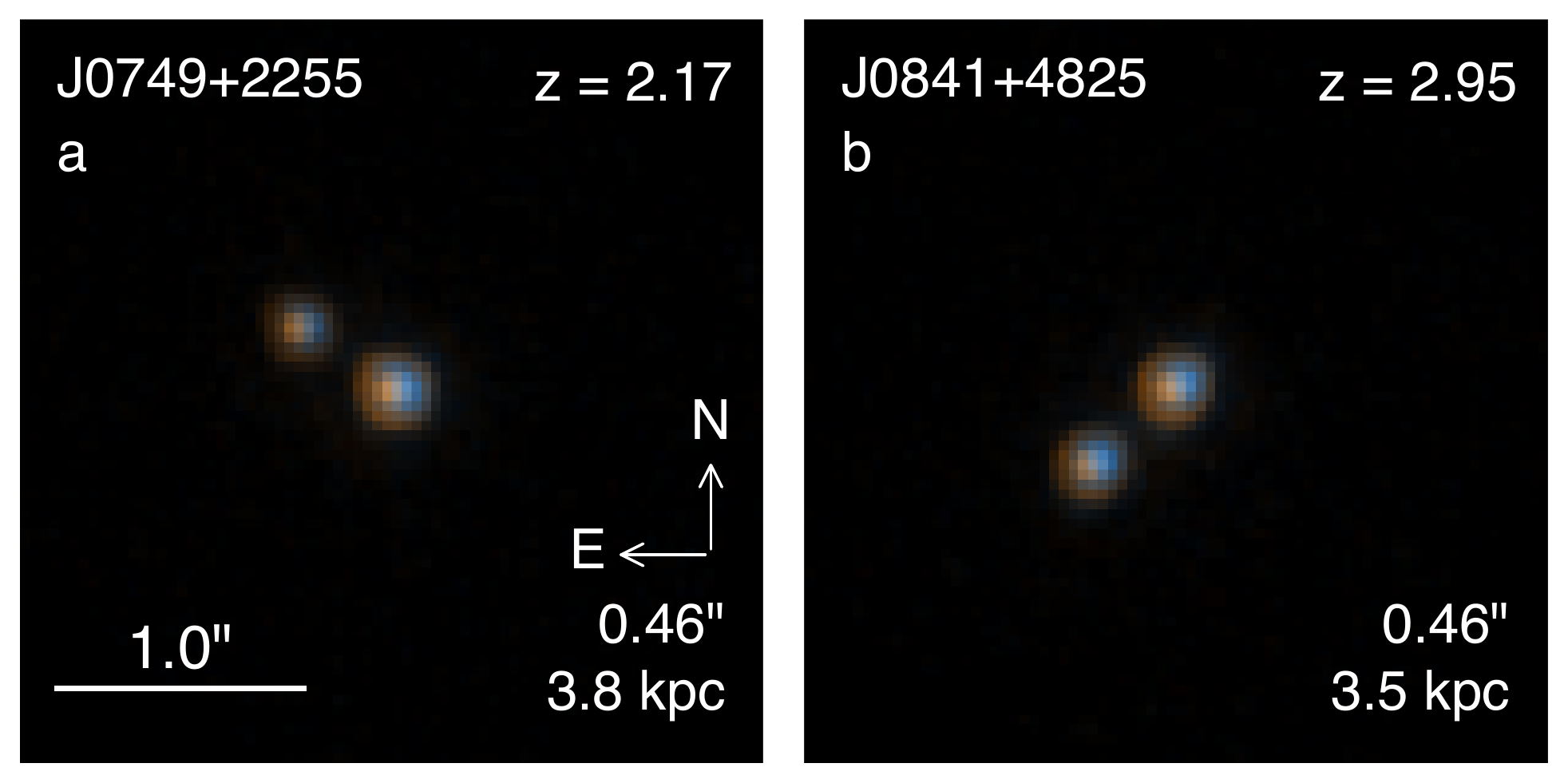}
\caption{\textbf{HST 2-band composite images of two double quasars.} The separations of resolved double cores in angular and physical units are marked. The double cores in J0749 (panel a) and J0841 (panel b) have similar colors, indicating both are most likely quasars, either from a physical pair, or multiple images of a gravitationally lensed quasar. } \label{fig:hst}
\end{figure}

\begin{figure}
\centering
\includegraphics[width=0.8\textwidth]{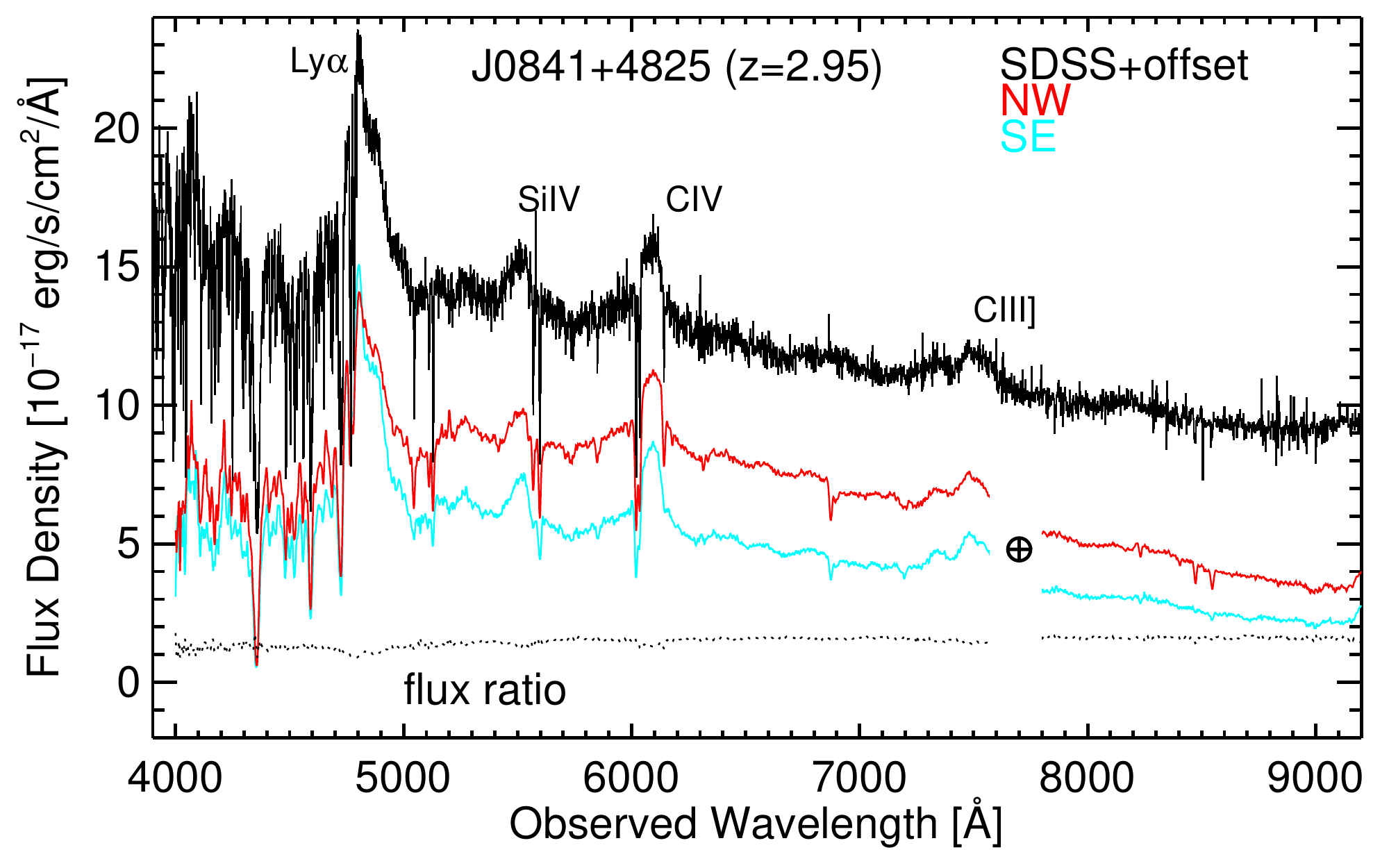}
\caption{\textbf{Spatially resolved optical spectroscopy of J0841.} The red and cyan lines are for the northwestern and southeastern cores, respectively. The gray dotted line denotes the flux ratio between the two cores. The circled-cross symbol denotes the telluric absorption region masked out in the red and cyan spectra. The black line is the spectrum from SDSS (with offset for clarity) that covers both cores. The two cores have similar spectral appearances, albeit with differences in the strengths of the broad emission lines (Methods). } \label{fig:spec}
\end{figure}

\begin{figure}
\centering
\includegraphics[width=0.6\textwidth]{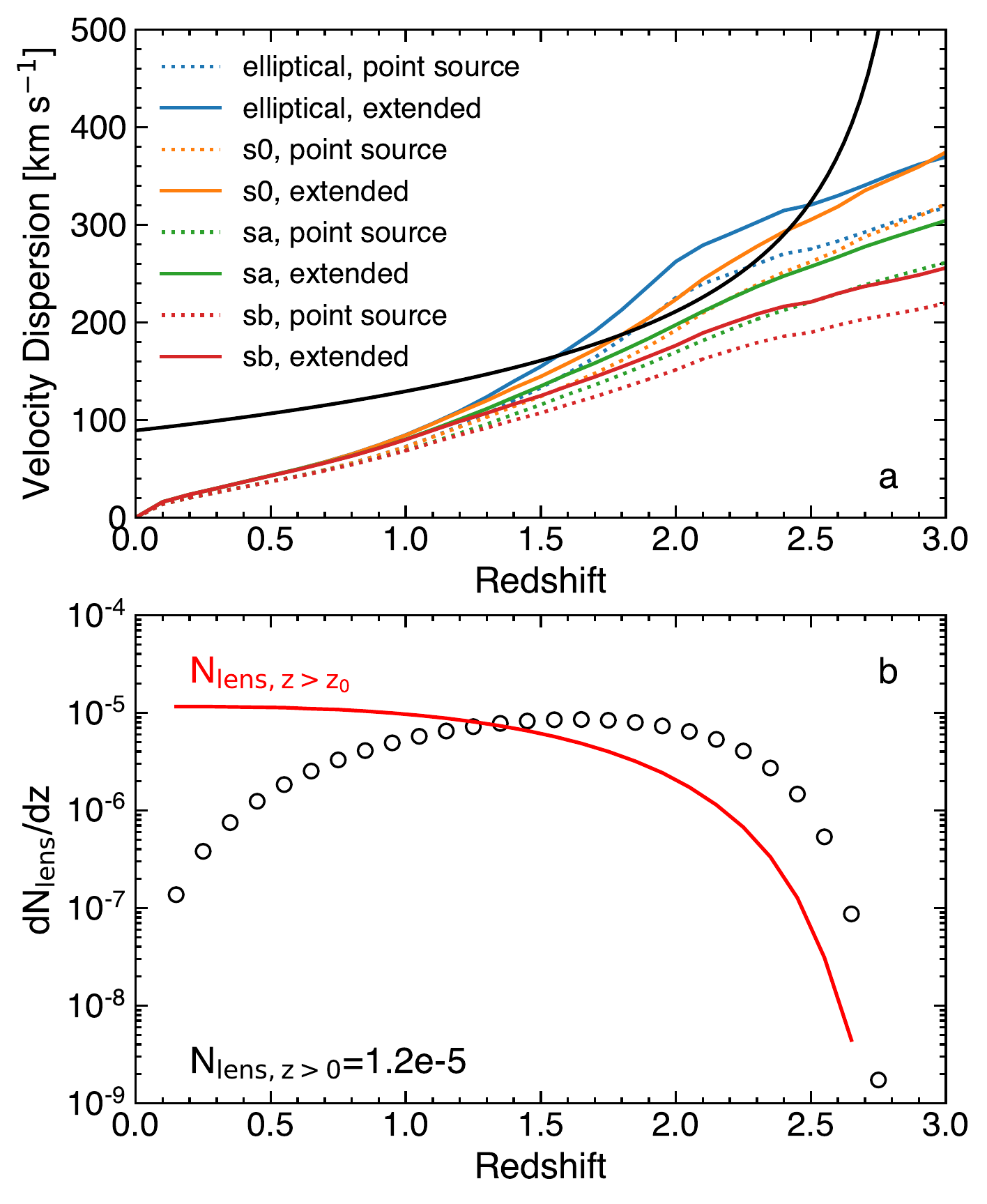}
\caption{\textbf{Lensing probability.} {\em Panel a:} the black line is the required stellar velocity dispersion for the lens at different redshift. The different colored lines are the $3\sigma$ upper limits from the non-detection of the lens galaxy in J0841 in HST F814W, for point sources (dotted lines) and extended sources (solid lines), respectively. We caution that our detection limits for extended sources may be overly optimistic, and more conservative estimates of the detection limits will lead to less stringent constraints on the lensing hypothesis (see Methods for details).  {\em Panel b:} the points are the differential number of lenses at different redshifts, estimated using the stellar velocity dispersion function and impact parameter for a hypothetical lensed quasar in J0841 (detailed in Methods). The red line is the cumulative number of lenses for lens redshift greater than the threshold. } \label{fig:lens}
\end{figure}

\clearpage

\noindent \textbf{\Large Methods}

\section{Varstrometry selection}

Our technique combines astrometry and variability in unresolved quasar doubles, referred to as ``varstrometry'' for convenience\cite{vodka1,vodka2}. Gaia measures source properties with a window size\cite{Gaia} of roughly $1\farcs2\times 2\farcs8$. Depending on the scanning direction of Gaia, it is possible that close pairs with sub-arcsec separations are still unresolved, and Gaia measures the system as a single point source. However, $\sim 80\%$ of the pairs with separations $\gtrsim 1\arcsec$ are resolved by Gaia DR2 [ref.~32]. Thus the varstrometry technique probes sub-arcsec separations in unresolved Gaia DR2 sources. In the Gaia astrometric solution, there is an extra noise term, astrometric\_excess\_noise, that describes the rms residuals that cannot be accounted for by measurement uncertainties. With zero parallax and proper motion in the astrometric solution, this quantity is then equivalent to the rms astrometric jitter defined in Eqn.\ (\ref{eqn:jitter}) for unresolved quasar pairs or lensed images. At low redshift, extended host galaxy light, and the random orientation of the galaxy relative to the Gaia scanning direction, may induce additional systematic errors in the photocenter measurements that can contribute to astrometric\_excess\_noise as well\cite{vodka1}.  

To select a clean sample to test the efficiency of the varstrometry technique in finding sub-arcsec double quasars, we start from the spectroscopically confirmed quasars from the SDSS DR7 and DR14 quasar catalogs\cite{Shen_etal_2011,dr14q}. We restrict to $z>2$ quasars so that host galaxy emission is negligible in the Gaia G band, and there is only a single Gaia match within a 3\arcsec\ radius to the SDSS position. We further limit to the brightest quasars with $G<19.1$ so that Gaia can measure photometry and astrometry to superb precision, resulting in 13,247 quasars. To ensure robust astrometric measurements, we further restrict to objects with Gaia visibility\_periods\_used$>8$, resulting in 11,537 objects. We then selected 15 of them with astrometric\_excess\_noise$>1.45$\,mas. We visually inspected the SDSS spectra of the 15 targets, and removed two objects with obvious stellar features (hence likely a quasar-star pair), ending up with 13 targets.

Since a small number of quasars with high astrometric\_excess\_noise may fall slightly below the $G=19.1$ flux limit, we imposed a secondary selection with astrometric\_excess\_noise$>3$\,mas and $G<19.5$. Only three additional targets meet these cuts, and one of them is removed with stellar features in the SDSS spectrum, leaving two additional targets. These two targets have nearly identical Gaia magnitudes ($G=19.3$) and astrometric\_excess\_noise. Our final target sample therefore includes 15 objects, and their distribution in the $G$ magnitude versus astrometric\_excess\_noise plane is shown in Extended Data Figure~1. For the parent sample statistics, however, we still use the sample of 11,537 objects with $G<19.1$ that dominates our target selection, since the selection to $G<19.5$ is highly incomplete. We note that one of the two confirmed double quasars, J0841, was from the ancillary selection. If we restrict our statistical constraints on the double-quasar frequency strictly to the main target selection, we have a slightly lower success rate, $>1/3$ instead of $>50\%$. The resulting double-quasar frequency among the parent quasar sample becomes slightly lower than $\sim 0.1\%$ but still roughly in line. Since our statistical estimates are all approximate, we ignore this detail in our following analysis.

The Gaia magnitude and astrometric\_excess\_noise cuts in our selection are imposed both to reduce potential contamination from systematics, and to yield a sample size reasonable for a HST program. If we were to lower these thresholds, we would have more candidates albeit with a potentially higher contamination rate. Since we have confirmed a very high success rate with our stringent Gaia cuts, the next step would be to lower the Gaia magnitude cut and astrometric\_excess\_noise cut to increase the target sample size, and to quantify the success rate at these less stringent Gaia cuts. 

Under these Gaia astrometric parameters and assuming a total fractional photometric variability of $10\%$, Eqn.~(\ref{eqn:jitter}) implies that our sample probes quasar doubles at separations between $\sim 30$\,mas and $\sim 1\arcsec$. At $z=2$, this angular separation range corresponds to projected separations between 250 parsec and 8.5 kpc, by far the most stringent limits that have been systematically explored\cite{Hennawi06,Hennawi10,Eftekharzadeh17} at $z>2$. But since SMBH pairs spend much longer time on kpc-scales than on smaller scales due to dynamical friction, we expect our methodology will preferentially discover kpc-scale double quasars. 

One caveat in the above derivation is that the mapping from astrometric jitter to physical separations is approximate at best. Because Gaia has not released time-resolved photocenter measurements and photometry, our analysis relied on published surrogates for the photometric variability and astrometric jitter. Additionally, the assumption that the fractional variability is similar for both components may break down. Therefore we do not necessarily expect a good agreement between the measured pair separation with HST and the predicted separation via Eqn.\ (1). It is also possible that a sub-arcsec pair, while fully within the window to measure astrometry, would suffer more from Gaia astrometric systematics than single sources. Nevertheless, targeting high-$z$ quasars with significant Gaia astrometric excess noise efficiently reveals systems with sub-arcsec companions that have compromised the Gaia astrometric solution, justifying this overall targeting strategy.  

\section{Follow up observations}

The 15 SDSS-Gaia quasars were then included in a HST Snapshot program to acquire two-band optical imaging in UVIS/F475W and F814W, with typical exposure times of 360/400\,sec in F475W/F814W. HST Snapshot programs observe submitted targets only if they fill the ``gaps'' between regularly scheduled programs. Four of the 15 targets were randomly observed, with three showing doubles as shown in Figure~1 and Extended Data Figure~2. The success rate is thus very high, demonstrating the efficiency of the varstrometry technique. The optical spectrum for J0841 is shown in Figure~2, and the SDSS spectra for the remaining three targets are shown in Extended Data Figure~3. Additional properties of the four targets are provided in Supplementary Table 1. 

Image decomposition of the HST data with GALFIT\cite{galfit} indicates all components in resolved doubles are consistent with point sources, with no evidence for extended host or additional components between resolved double cores. For the three resolved double cores, we measure a flux ratio (in F475W/F814W) of 4.4/3.6 (J0749), 1.3/1.7 (J0841) and 46/4.5 (J0905), with $1\sigma$ fractional uncertainties of $<10\%$. The GALFIT results are shown in Extended Data Figure~5.

We obtained Gemini Directors Discretionary Time for one of the three HST-resolved doubles, J0841. This particular target was chosen due to visibility constraints and the limited amount of DDT time available for the highest image quality (the top 20\%), which is necessary to spatially resolve the double from ground-based seeing-limited observations. Gemini fast-turnaround and regular proposals were also submitted for the other targets but were unsuccessful due to the challenges to secure the required highest image-quality time. With the approved DDT time, J0841 was observed on 21 May 2020 UT with optical slit spectroscopy with the GMOS spectrograph on Gemini-North (Program GN-2020A-DD-106). We chose optical spectroscopy over near-IR spectroscopy to ensure detection of both cores in this high-redshift quasar. But future near-IR slit spectroscopy to spatially resolve the narrow emission lines is warranted. The observations were carried out during superb conditions, with seeing FWHM$\sim$0\farcs4--0\farcs5, sufficient to resolve the double.  We adopted the longslit with the R150 grating and a 0\farcs75 slit width. It offers a spectral resolution of $R\sim420$ (corresponding to $\sigma_{{\rm inst}}\sim$210 km s$^{-1}$) spanning the observed $\sim$400--950 nm with a pixel scale of 1.93 {\rm \AA} pixel$^{-1}$ along the wavelength direction. The slit was placed across the two cores along PA=132.0$^{\circ}$. The total exposure time was 1,848 s, which was divided into two individual exposures dithered at two central wavelengths (700 nm and 720 nm) to cover the CCD chip gaps and to help reject cosmic rays. We reduced the data following standard IRAF procedures\cite{Tody1986} using the PyRAF package [http://ast.noao.edu/sites/default/files/GMOS\_Cookbook/Processing/PyrafProcLS.html]. The 2D spectra were wavelength calibrated using CuAr lamp lines and stored in vacuum wavelength. The white dwarf EG 131 was observed as the standard star for flux calibration during the same night.

To spatially resolve the double cores, we first collapse the 2D spectrum along the wavelength direction to construct a spatial profile (see Extended Data Figure~6). For J0841 at $z=2.95$, the optical emission is dominated by the accretion disk and broad emission lines, and thus both cores can be treated as point sources, with the same point-spread-function (PSF) determined by seeing. The two cores are clearly resolved in the 2D spectrum, and the wavelength-collapsed profile provides a robust measure of the seeing profile. We found a single Gaussian fits the PSF core well, with a measured FWHM of 0\farcs4 (consistent with the seeing at the time of the observations) and a spatial offset of 5.8 pixels (or $\approx 0\farcs47$) between the two cores, consistent with the separation measured from HST imaging. With the measured PSF, we then decompose the two cores at each wavelength to extract the 1D spectra for both cores. We use a single Gaussian for each component, with the separation and Gaussian dispersion fixed from the spatial profile derived earlier, but allowing the amplitude of each Gaussian to vary in case the two components have different spectral shapes. The spatially-decomposed 1D spectra for the two cores are shown in Figure~2. 

The two spectra are very similar in appearances with nearly identical redshifts, except for subtle differences in spectral shape, strengths of certain intervening absorption lines, and the strengths of broad emission lines (Extended Data Figure~7). While these spectral differences in general can be explained in the lensing scenario, the different flux ratios in the continuum and in the broad emission lines require a microlensing event\cite{Sluse}, thus reducing the lensing probability. A simpler explanation for the spectral dissimilarity is that J0841 is a physical quasar pair. Assuming the two quasars have the same BH mass, the factor of 1.5 difference in continuum luminosity will not reflect in their broad-line width difference, since broad-line width scales as $L^{1/4}$ if the broad-line region is virialized\cite{Shen13}. Bearing in mind the large ($\sim 0.4$\,dex) systematic uncertainties, we estimate BH masses for the two quasars in J0841 using the single-epoch virial mass method\cite{Shen13} as $10^{9.4\pm0.4}\,M_\odot$ (SE) and $10^{9.6\pm0.4}\,M_\odot$ (NW), where the spectral fits were performed using the public \texttt{QSOFIT} code\cite{Shen19}, and the mass errors represent 1$\sigma$ standard deviations and are dominated by systematic uncertainties.  

For J0749, while we have not obtained spatially-resolved spectroscopy, a 15~GHz (2~cm wavelength) image obtained with the Very Long Baseline Array (VLBA) shows both cores (Extended Data Figure~4), confirming its double-quasar nature.  These two quasars must be at the same redshift given the lack of multiple emission-line systems in the SDSS spectrum. The full analysis of these VLBA data (including data in additional frequency bands) will be presented elsewhere.

\section{Lensing model of J0841}

We perform simple lens modeling using a singular isothermal sphere (SIS), shown to be a good approximation for sub-arcsec lensing\cite{Oguri06}. The SIS model constrains the velocity dispersion of the lens galaxy as a function of the unknown lens redshift, as shown by the black line in Figure~3 (panel a). From the HST non-detection of the lens (with exposure time of 400 sec in F814W), we constrain the $3\sigma$ upper limit of the galaxy luminosity for a range of morphological types with different spectral shapes\cite{Kinney1996}. For each galaxy type, we consider both a point source and an extended source. For the point source case, the $3\sigma$ optimal extraction corresponds to a limiting AB magnitude of $27$ in F814W. For the extended source case, we assume all light is uniformly distributed within a 0\farcs2 radius, i.e., filling the area between the two cores, reaching a limiting AB surface brightness of 24.1\,${\rm mag\,arcsec^{-2}}$. We then convert the galaxy luminosity to velocity dispersion using empirical relations measured for local galaxies\cite{Bernardi}. The scatter in velocity dispersion at fixed luminosity is $\lesssim 25\%$ and we assume no evolution in these scaling relations.  

We note that the limits for the extended source case are optimistic, since the lens galaxy can be more extended and blend into the PSF wings of the quasars. For this reason, we proceed with caution on our lensing probability assessment. This exercise is presented as a general methodology to test the lensing hypothesis, even though the current data are insufficient to draw definitive conclusions. 

The upper limits on the velocity dispersion inferred from the HST non-detection are plotted against redshift in Figure~3 (panel a) for different galaxy types. While the HST imaging is not deep enough, these upper limits meaningfully rule out a lens at $z<1.5$. Lenses at higher redshifts are still possible, considering the uncertainties in the estimated upper limits from HST non-detection. To further evaluate this possibility, we estimate the number of galaxies with sufficient velocity dispersion along the line-of-sight towards J0814 from random superposition. We use the stellar velocity dispersion function measured for local early-type galaxies\cite{Sheth}, which only shows mild evolution\cite{Bezanson} towards $z\sim 2$. In the case of an SIS lens, the flux ratio of two images is given by
$r = (1 + y) / (1 - y)$, where y is the angular separation between the source and the lens center normalized by the Einstein radius.
For the flux ratio of $r = 1.5$, we have $y = 0.2$, so the impact parameter required to reproduce the observed flux ratio is
0\farcs046. Our selection strategy is able to randomly discover lensed images with separations within a factor of two of the observed $\sim 0\farcs4$ separation. The lower limit is set by the HST resolution and the upper limit is set by Gaia, i.e., larger separations will be resolved by Gaia. This image separation range corresponds to a range of SIS velocity dispersions that enclose a factor of $\sqrt{2}$ around the fiducial value, which we integrate over to obtain the lens galaxy number density at given lens redshift. The cumulative number of random intercepting galaxies at redshifts greater than a threshold is shown as the red line in Figure~3 (panel b). The expected number is thus $\sim 6\times10^{-6}$ for a $z\gtrsim 1.5$ lens. If we adopt a more conservative sensitivity (e.g., reducing by one magnitude) of the lens galaxy in the extended-source case, the lensing modeling can rule out a lens galaxy at $z>1.2$ (as opposed to $z>1.5$). The corresponding cumulative lens probability would increase from $6\times10^{-6}$ to $8\times10^{-6}$.

We further estimate the expected total number of sub-arcsec doubly lensed quasars with comparable image flux ratios, for the parent population of $z>2$ quasars. Since the lensing probability increases towards higher source redshift, we can conservatively use the single-lensing probability estimated above for J0841 at $z\sim 3$. There are $\sim 11,000$ parent $z>2$ quasars. To estimate the boost factor in source number density due to lensing magnification bias\cite{Oguri19}, we use the latest optical quasar luminosity function measured\cite{Ross13} at $z>2$ to derive a boost factor of $\sim 30$. Therefore we expect $\sim 2-3$ doubly lensed quasars out of the $\sim 11,000$ parent quasars. Our program revealed that at least half of the 15 quasars with the highest varstrometry signals should be sub-arcsec double quasars, already exceeding the lensing expectation with this subsample of 15 objects alone. Considering our selection is by no means complete, the general population of sub-arcsec double quasars at $z>2$ may be too abundant to be explained by the lensed population. However, as mentioned earlier, the uncertainties in the non-detection limits make it difficult to rule out the possibility that these double quasars are dominated by the lensing population. Deeper limits (preferentially from infrared imaging) are required to improve this statistical constraint.

We now roughly estimate the probability that J0841 is an expected lensed quasar from our lensing probability analysis. While the efficiency of our varstrometry selection of high-redshift quasar doubles is high ($\gtrsim 50\%$), we cannot formally estimate the completeness of the selection technique due to the lack of detailed knowledge on Gaia time series data, astrometric modeling, the unknown distributions of pair separations and variability amplitudes. Nevertheless, we do not expect that the completeness is very high, since the technique is likely to miss most genuine sub-arcsec pairs with smaller astrometric jitter. We do not expect the completeness to be very low either ($<10\%$), otherwise there may be an overabundance of sub-kpc binary SMBHs. Thus we assume a reasonable completeness of 50\% of our selection, which means the 15 HST targets should include one sub-arcsec lensed quasar. The probability that it happens to be observed in the 4 randomly observed targets is 27\% (i.e., $1-C_{14}^{4}/C_{15}^4$). Thus the total probability of J0841 being a lensed quasar is $0.27$. The conditioned probability that J0841 is a lensed quasar but also at $z>2.8$ is further reduced to $0.27(2242/11537)\approx 0.05$, where 2242 is the number of $z>2.8$ quasars in the parent sample. So there is still a small probability of J0841 being a lensed quasar. Since we have likely underestimated the total number of lensed quasars in our sample, the actual probability of J0841 being a lensed quasar could be significantly higher. But deep IR imaging is required to confirm the required lens in J0841.

\section{Implications from small-scale quasar clustering}

The expected physical pair frequency can be estimated from small-scale quasar clustering measurements\cite{Eftekharzadeh17,Kayo}.  Although there is no such clustering measurement on kpc scales, we use existing measurements at $z<2$ and larger scales to derive some crude estimates. The small-scale quasar pair sample in \cite{Kayo} enabled the measurement of the quasar correlation function on $\gtrsim 15\, {\rm kpc}$ projected scales. Below $\sim 50\,{\rm kpc}$, there is evidence that the 3D correlation function steepens to a $r^{-3}$ power law, which would imply a constant pair fraction for each decade in scale. Based on the pair statistics from that work and assuming their pair sample is complete, the extrapolated pair fraction on kpc-scale is then $\sim 0.05\%$. A somewhat higher pair fraction of $\sim 0.26\pm0.18\%$ was recently reported based on ground-based imaging and spectroscopic follow-up\cite{Silverman20}. Furthermore, it is possible that the abundance of kpc-scale quasar pairs is much higher at $z>2$ given the increased specific merger rate and the enhanced probability that both SMBHs are quasars due to merger-induced fueling. This possibility is particularly relevant for the luminous quasars at $z>2$ as in our bright Gaia sample. Therefore we may expect a much higher kpc-scale quasar pair fraction at $z>2$, which will be tested with our systematic search using varstrometry.

\section{J0905 as a star-quasar superposition}

The color discrepancy in the double cores in J0905 suggests that the fainter and redder component is a star. The measured F475W-F814W color for the redder component is consistent with a M2V star. We add a M2V star spectrum to the SDSS quasar spectrum of J0905 by fixing the flux ratio in the F814W band. The predicted flux ratio in the F475W band is 0.02, consistent with that measured from HST data, and the predicted flux ratio in Gaia G band is 0.1. The stellar absorption features of the star are difficult to identify given the flux errors. Therefore we could not have rejected J0905 as a quasar-star superposition from the SDSS spectrum.

\noindent\textbf{Data availability} The SDSS spectrum for J0841 is publicly available at https://www.sdss.org/. The HST data are publicly available via the Mikulski Archive for Space Telescopes (MAST) at https://archive.stsci.edu with program number HST-GO-15900. The raw data for the Gemini spectrum are publicly available at https://archive.gemini.edu/ with program ID GN-2020A-DD-106, and the reduced spectrum is provided in the data for Figure~2. The catalog data for parent SDSS quasars are available in \cite{Shen_etal_2011,dr14q}, and the astrometric data are publicly available from Gaia Data Release 2 at https://gea.esac.esa.int/archive/. Additional data (preliminary VLBA images) that support the plots within this paper and other findings of this study are available from the corresponding author upon reasonable request.

\noindent\textbf{Code availability} The code (GALFIT) used to decompose the HST images is publicly available at https://users.obs.carnegiescience.edu/peng/work/galfit/galfit.html.

\begin{figure*}
\centering
\renewcommand\thefigure{E1}
\includegraphics[width=0.8\textwidth]{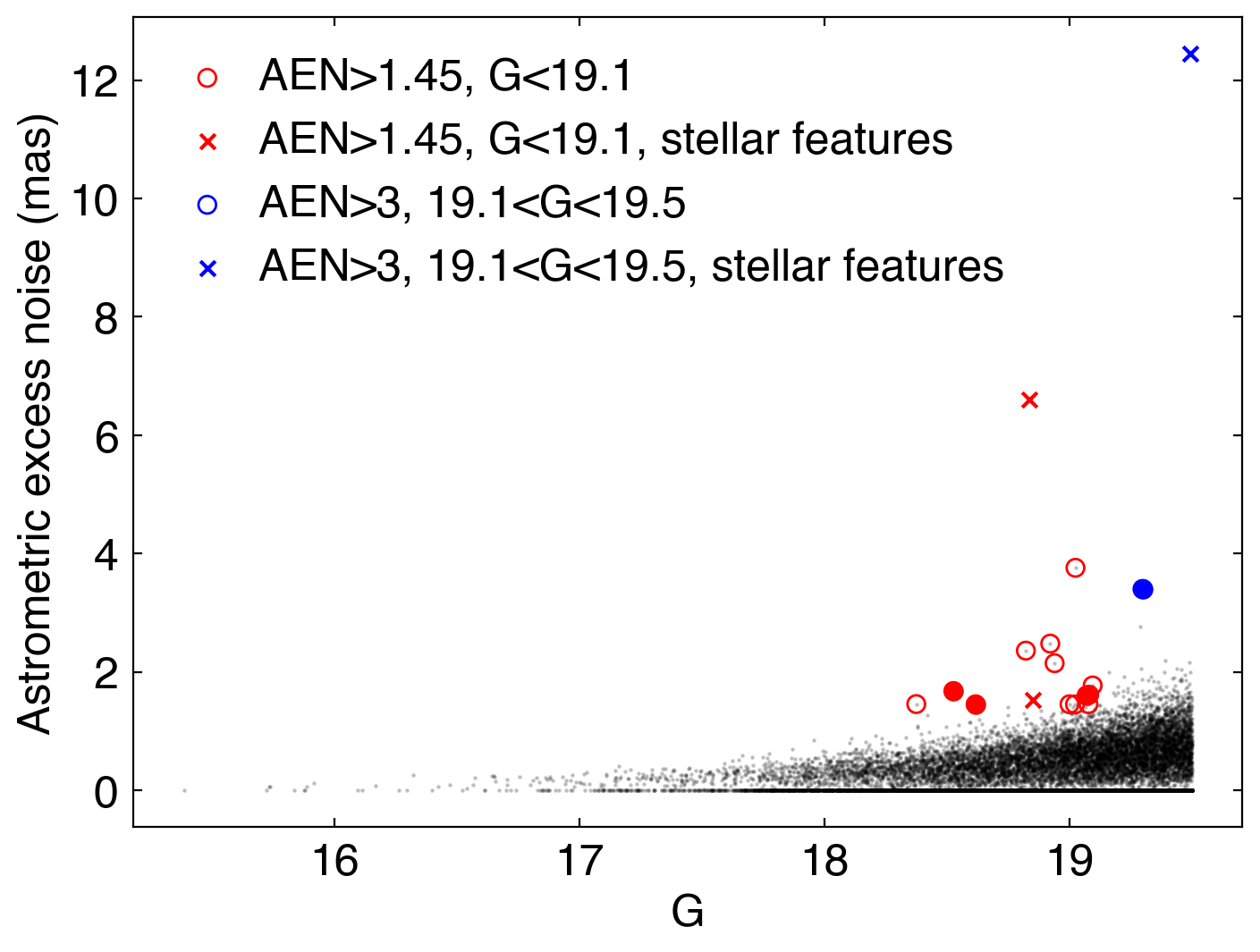}
\caption{\textbf{Gaia selection of candidate double quasars.} The black dots are the parent $z>2$ SDSS quasars with single Gaia matches. The red and cyan circles are our final sample of 15 targets (there are two objects in cyan circles that overlap with each other), and the crosses are excluded objects based on apparent spectral features that indicate a star-quasar superposition (see details of target selection in Methods). Most of the objects at $G>19.1$ with high astrometric excess noise are excluded from our extended target selection (the blue circles). The four targets that have been followed up by HST are indicated by filled circles, and the remaining targets are indicated by open circles. } \label{fig:gaia_selection}
\end{figure*}

\begin{figure*}
\centering
\renewcommand\thefigure{E2}
\includegraphics[width=0.8\textwidth]{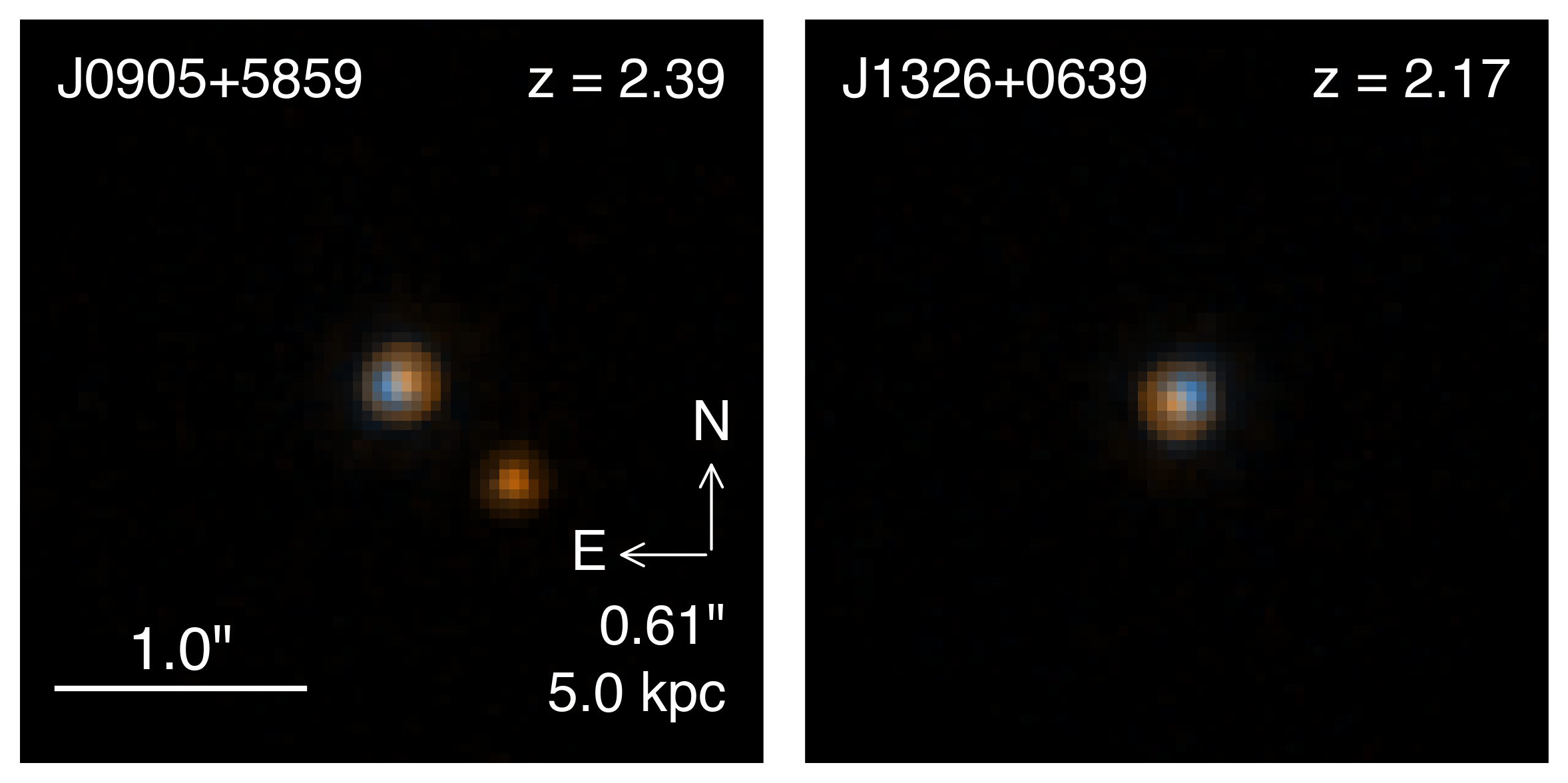}
\caption{\textbf{HST 2-band composite images of the remaining two targets.} The different colors in the two cores in J0905 suggest it is a quasar-star superposition, and J1326 is not resolved at the HST resolution of $\sim 0.1\arcsec$. } \label{fig:nonqso}
\end{figure*}

\begin{figure*}
\centering
\renewcommand\thefigure{E3}
\includegraphics[width=0.8\textwidth]{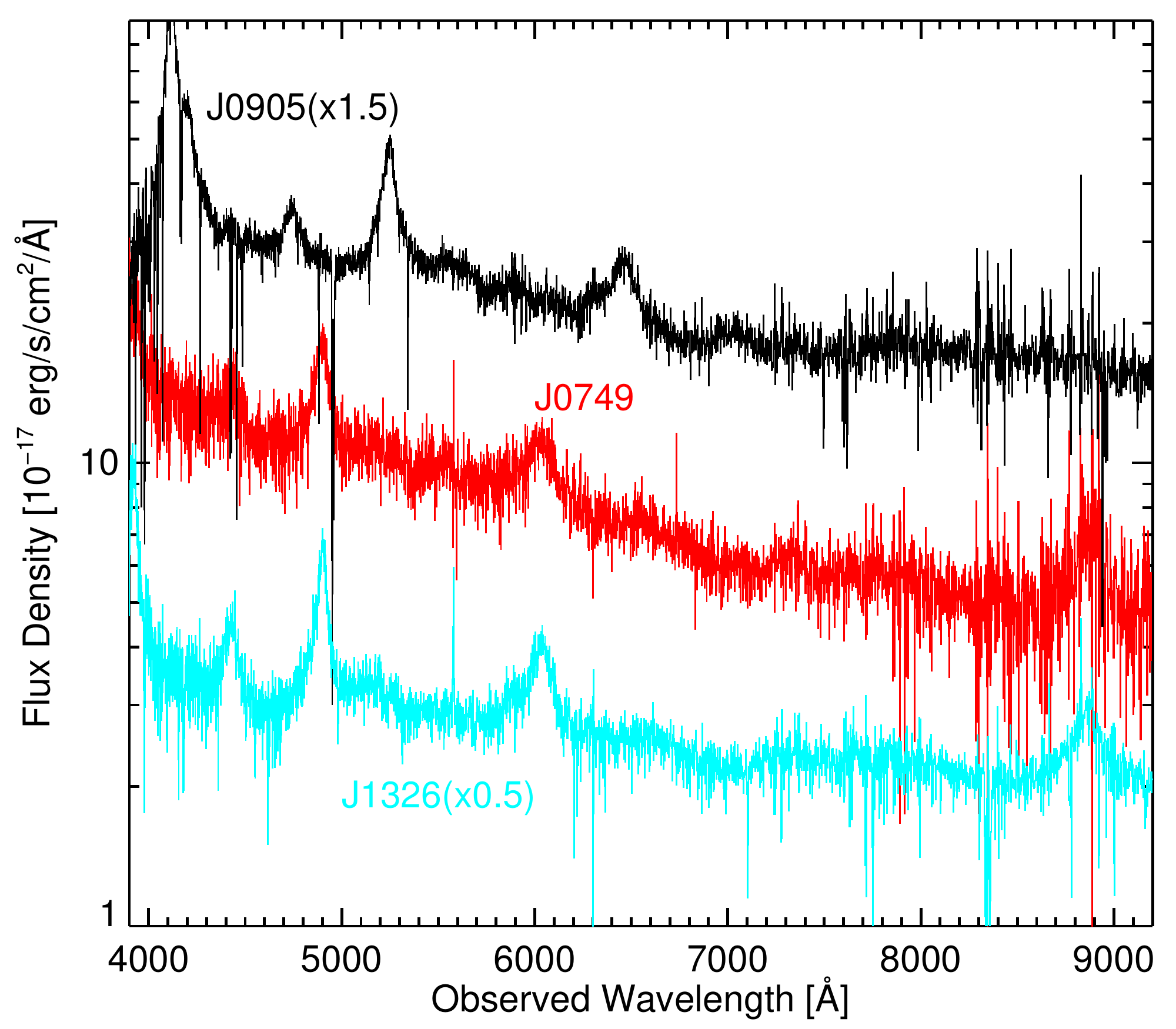}
\caption{\textbf{SDSS spectra for three targets observed by HST.} The flux densities of J0905 and J1326 have been scaled for the clarity of the figure. } \label{fig:sdss_spec}
\end{figure*}

\begin{figure*}
\centering
\renewcommand\thefigure{E4}
\includegraphics[width=0.98\textwidth]{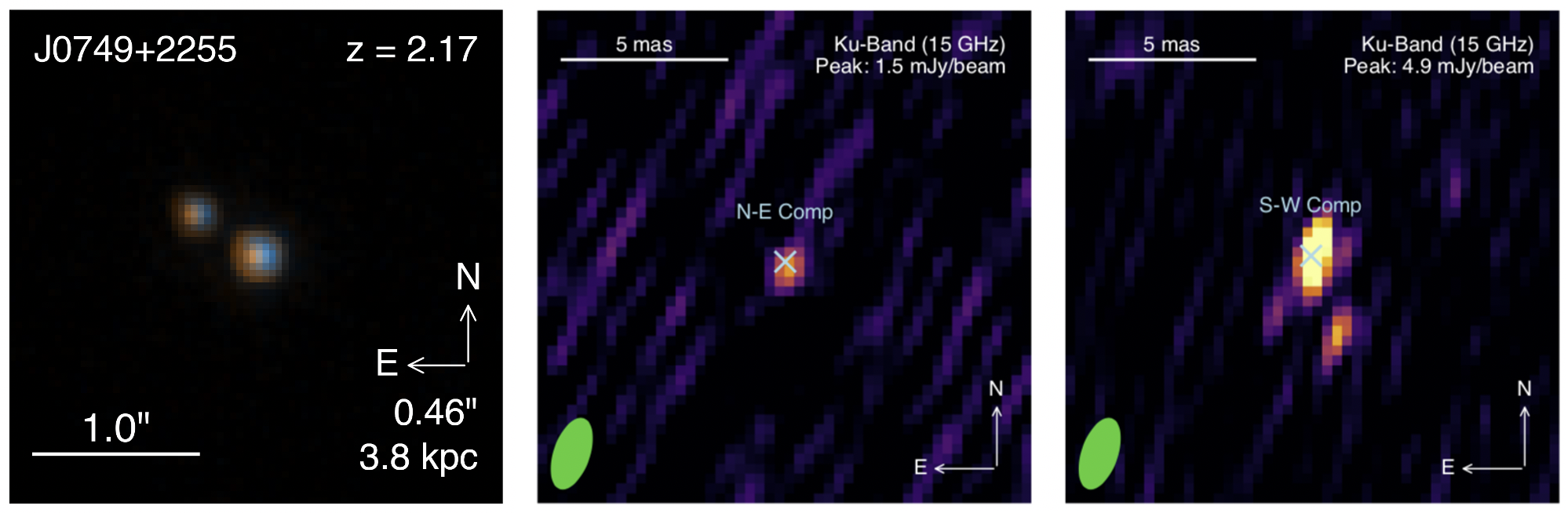}
\caption{\textbf{Preliminary analysis of VLBA imaging for J0749.} Panel a shows the optical HST image and panels b, c show the VLBA detection of both optical cores in Ku-band (15\,GHz) with few-milliarcsec (mas) resolution, indicating both cores are quasars. There are also morphological differences in the radio images of the two cores. The VLBA beam is shown in the bottom left corner with the green ellipse. } \label{fig:J0749_vlba}
\end{figure*}

\begin{figure*}
\centering
\renewcommand\thefigure{E5}
\includegraphics[width=0.45\textwidth]{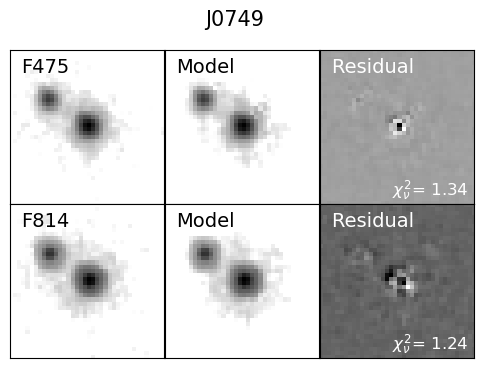}
\includegraphics[width=0.45\textwidth]{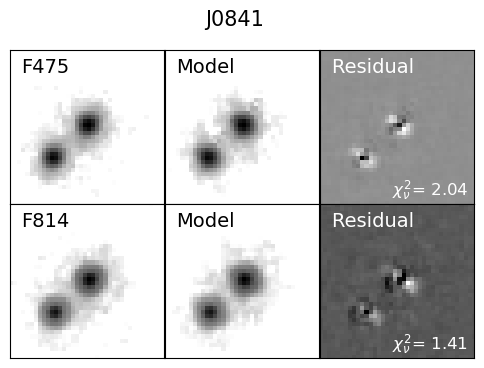}
\includegraphics[width=0.45\textwidth]{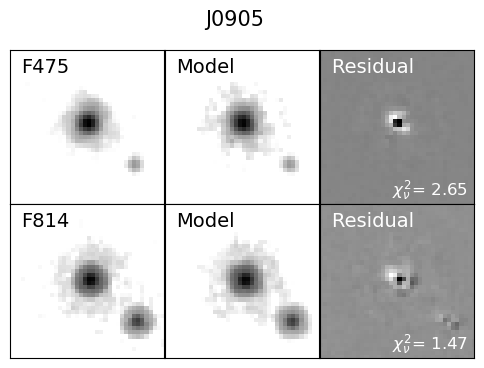}
\includegraphics[width=0.45\textwidth]{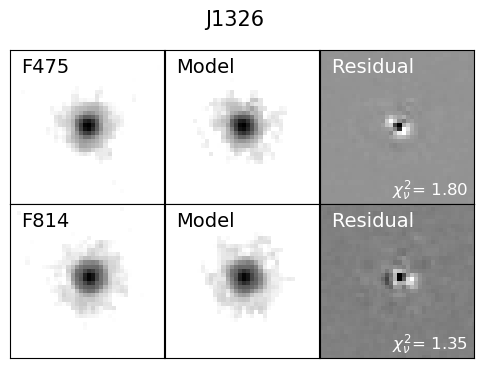}
\caption{\textbf{HST image decomposition.} For each of the four systems shown in panels a-d, we show the original, model and residual (model$-$original) images, where the models are constructed with GALFIT. The reduced $\chi^2$ values are provided in the lower right corner of each row.} \label{fig:galfit}
\end{figure*}

\begin{figure*}
\renewcommand\thefigure{E6}
\includegraphics[width=0.6\textwidth]{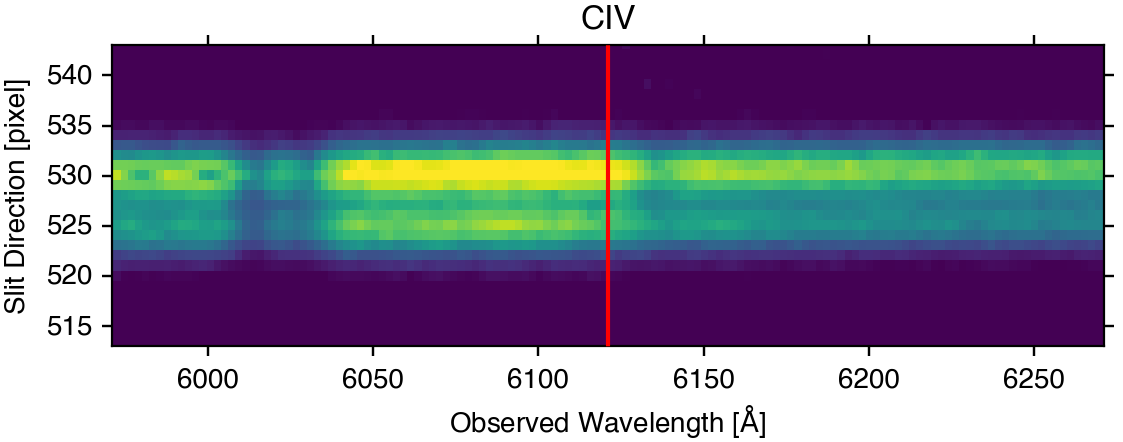}\quad
\includegraphics[width=0.3\textwidth]{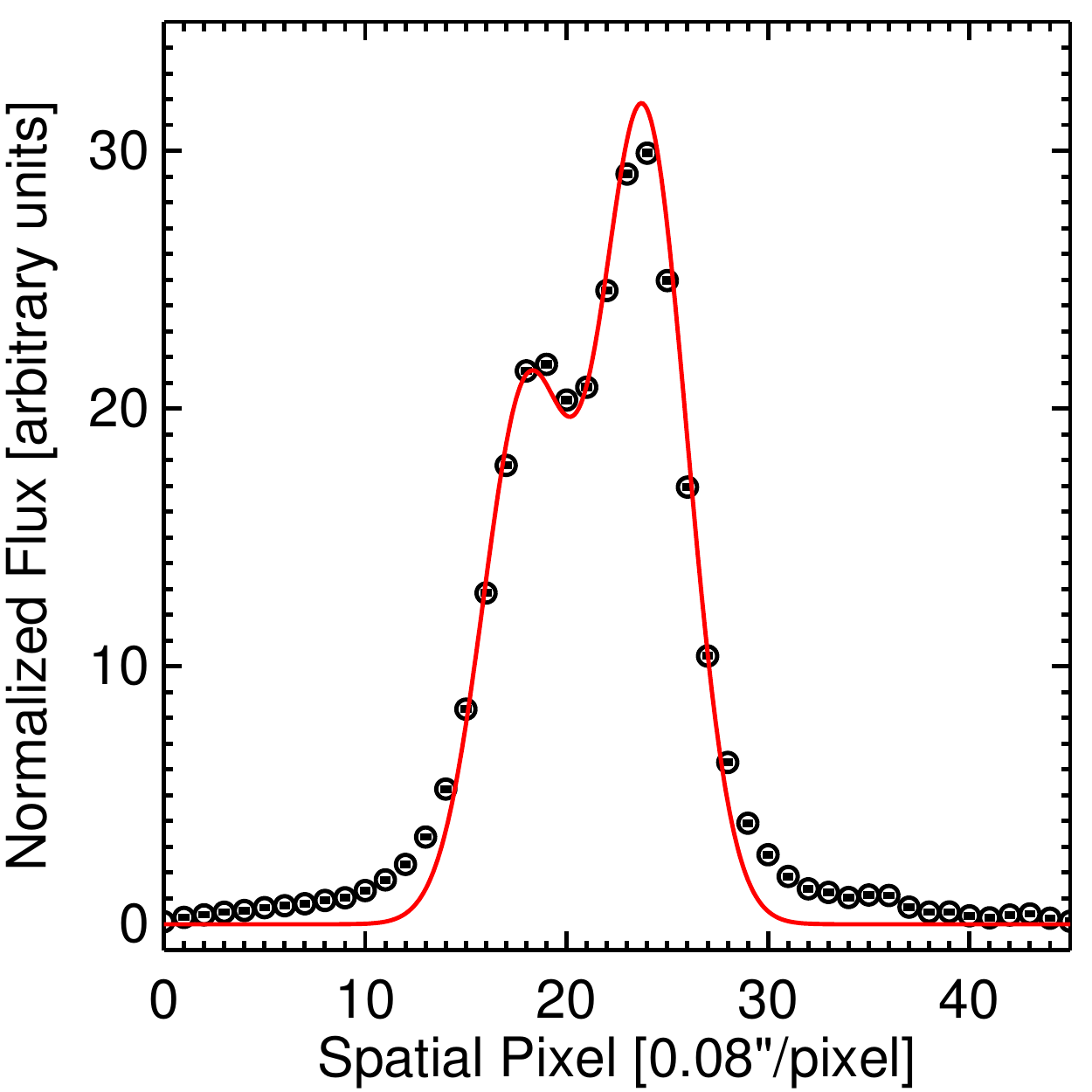}
\caption{\textbf{2D spatial profile of J0841 in slit spectroscopy.} {\em Left:} the two-dimensional spectrum around the \CIV\ line, where two sources separated by $\sim 0\farcs46$ ($\sim 5.8$\,pixels) are visible. {\em Right:} spatial profile of the wavelength-collapsed spectrum, which we used to measure the separation between the two sources in the slit spectrum. The points are the pixel data plotted with 1$\sigma$ standard deviation error bars (the errors are very small), and the red line is a double-Gaussian model.} \label{fig:2dspec}
\end{figure*}

\begin{figure*}
\centering
\renewcommand\thefigure{E7}
\includegraphics[width=0.8\textwidth]{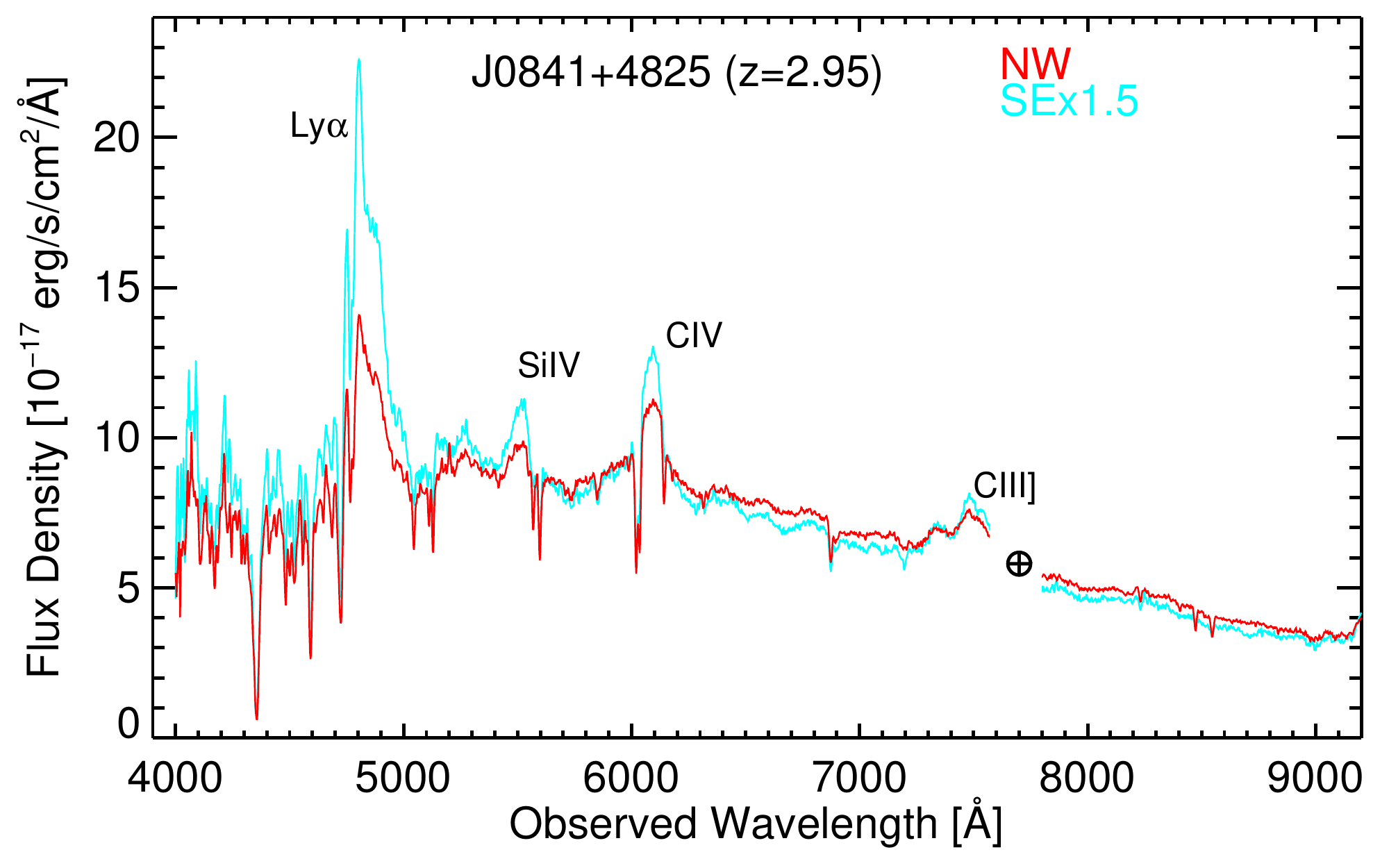}
\caption{\textbf{Spectral comparison of the two cores in J0841.} The red and cyan lines are for the northwestern (NW) core and the southeastern (SE) core, respectively. The circled-cross symbol denotes the telluric absorption region masked out in the spectra. There are notable differences in the strengths of the broad emission lines after the flux of the SE core is scaled to roughly match the continuum flux of the NW core. There are also slight differences in the continuum shape and strengths in certain intervening absorption lines. } \label{fig:speccomp}
\end{figure*}


\clearpage
\noindent\textbf{\huge Supplementary Information}

The target coordinates, Gaia DR2 parameters, separations and flux ratios measured from HST imaging are provided in Supplementary Table \ref{tab:summary}. We also estimate the expected pair separations, using surrogates of astrometric jitter and photometric variability from Gaia DR2 cataloged parameters. The astrometric jitter is estimated using the quadratic sum of Gaia measured parallax, proper motion$\times$time, and astrometric excess noise. The total fractional photometric variability is estimated using Gaia reported photometric uncertainties and the number of observing epochs\cite{vodka1} since Gaia DR2 did not publish photometric variability. 

For J0749 and J0841, both components are quasars and we assume the same fractional photometric variability. Using the average flux ratio in the HST bands, and Eqn.~3 in \cite{vodka1}, we estimate the expected pair separations. For J0905, the companion is a non-variable star, so we used Eqn.~5 in \cite{vodka1} to estimate the pair separation, using a flux ratio of 0.1 in Gaia G band from our modeling in Methods. J1326 does not have flux ratio measurement, so we used Eqn.~2 in \cite{vodka1} assuming that both (unresolved) components are quasars with equal mean fluxes. Given the assumptions in these estimations and the use of surrogates from Gaia DR2, the agreement between the expected and observed pair separations is not great, but consistent in the ballpark. 

\newgeometry{margin=1cm}
\begin{landscape}
\begin{table}
\caption {Summary of Targets} \label{tab:summary}
\begin{center}
\begin{tabular}{cc c cccccc} 
\hline \hline
Target & $z$ & $\sigma_{\rm AEN}$ & $\sigma_{\rm photo}$ & parallax & PM (RA/DEC) & flux ratio & $D_{\rm obs}$ & $D_{\rm exp}$ \\ 
 & & [mas] &  & [mas] & [mas/yr] &  &  [\arcsec] & [\arcsec] \\
\hline
J074922.97$+$225511.8 & 2.17 & 1.452 & 0.14 & $-0.38\pm 0.40$ & $-0.93\pm0.64$/$1.91\pm0.37$ & 4.4 (F475W) & 0.46 & 0.11 \\
             & & & & & & 3.6 (F814W) &  & \\
J084129.77$+$482548.3 & 2.95 & 3.399 & 0.09 & $-1.49\pm 1.02$ & $-3.68\pm1.17$/$8.54\pm1.33$ & 1.3 (F475W) & 0.46 & 0.41 \\
             & & & & & & 1.7 (F814W) & & \\
J090501.12$+$585902.3 & 2.39 & 1.676 & 0.08 & $-0.29\pm0.34$ & $0.03\pm0.48$/$-0.52\pm0.48$ & 46 (F475W) & 0.61 & 0.27 \\
             & & & & & & 4.5 (F814W) & & \\
J132614.08$+$063909.9 & 2.17 & 1.604 & 0.03 & $-0.18\pm0.55$ & $1.07\pm0.98$/$0.71\pm0.48$ & N.A. & $<0.1$ & 0.19 \\
\hline
\end{tabular}\\
Notes: The astrometric jitter is estimated by the quadratic sum of astrometric excess noise ($\sigma_{\rm AEN}$), parallax and proper motion distance over 22 months (observing period of Gaia DR2). The expected pair separation $D_{\rm exp}$ is estimated using the astrometric jitter, Gaia fractional photometric RMS variability ($\sigma_{\rm photo}$) and mean flux ratio. For J0749 and J0841 we use Eqn.~3 in \cite{vodka1}; for J0905 we use Eqn.~5 in \cite{vodka1} since the companion is a non-variable star; for J1326 we use Eqn.~2 in \cite{vodka1} since the flux ratio is unknown. All quoted error bars are $1\sigma$ standard deviations. 
\end{center}
\end{table}
\end{landscape}
\restoregeometry

\end{document}